\documentclass[journal=jacs,manuscript=communication]{achemso}

\title{Mixed-valence compounds as polarizing agents for Overhauser dynamic nuclear polarization in solids}

\author{Andrei Gurinov}
\affiliation{NMR Spectroscopy group, Bijvoet Center for Biomolecular Research, Utrecht University, Padualaan 8, 3584, CH Utrecht, The Netherlands}
\author{Benedikt Sieland}
\affiliation{Department of Chemistry, Paderborn University, Warburger Strasse 100, Paderborn, D‐33098 Germany}
\author{Andrei Kuzhelev}
\affiliation{Goethe University Frankfurt am Main, Institute of Physical and Theoretical Chemistry, Center for Biomolecular Magnetic Resonance, Max von Laue Str. 7, 60438 Frankfurt am Main, Germany}
\author{Hossam Elgabarty}
\affiliation{Dynamics of Condensed Matter and Center for Sustainable Systems Design, Chair of Theoretical Chemistry, University of Paderborn, Warburger Str. 100, Paderborn, Germany}
\author{Thomas D. K\"uhne}
\affiliation{Dynamics of Condensed Matter and Center for Sustainable Systems Design, Chair of Theoretical Chemistry, University of Paderborn, Warburger Str. 100, Paderborn, Germany}
\author{Thomas Prisner}
\affiliation{Goethe University Frankfurt am Main, Institute of Physical and Theoretical Chemistry, Center for Biomolecular Magnetic Resonance, Max von Laue Str. 7, 60438 Frankfurt am Main, Germany}
\author{Jan Paradies}
\affiliation{Department of Chemistry, Paderborn University, Warburger Strasse 100, Paderborn, D‐33098 Germany}
\author{Marc Baldus}
\affiliation{NMR Spectroscopy group, Bijvoet Center for Biomolecular Research, Utrecht University, Padualaan 8, 3584, CH Utrecht, The Netherlands}
\author{Konstantin L. Ivanov}
\affiliation{International Tomography Center, Siberian Branch of the Russian Academy of Sciences, Novosibirsk, 630090, Russia and Novosibirsk State University, Novosibirsk, 630090, Russia}
\author{Svetlana Pylaeva}
\affiliation{Dynamics of Condensed Matter and Center for Sustainable Systems Design, Chair of Theoretical Chemistry, University of Paderborn, Warburger Str. 100, Paderborn, Germany}
\email{svetlana.pylaeva@upb.de}

\abbreviations{IR,NMR,UV}

\begin{document}

\begin{abstract}
In this paper we investigate a novel set of polarizing agents -- mixed-valence compounds -- by electron paramagnetic resonance methods and demonstrate their performance in high-field Dynamic Nuclear Polarization (DNP) experiments in the solid state. Mixed-valence compounds constitute a group of molecules, in which molecular mobility persists even in solids. Consequently, such polarizing agents can be used to perform Overhauser-DNP experiments in solid-state, with favorable conditions for dynamic nuclear polarization formation at ultra-high magnetic fields.\end{abstract}

{\bf Introduction}

Dynamic Nuclear Polarization (DNP)\cite{MalyRev08,CorziliusRev2020} has become a widely used method for signal enhancement in various Nuclear Magnetic Resonance (NMR) experiments. DNP NMR has allowed applications that were not deemed feasible before: from proteins in cells,\cite{SidInCell} to atomistic studies of mesoporous materials,\cite{SENSDNP,CatalysisDNP} and to clinical applications of dissolution DNP.\cite{ARDENKJAERLARSEN20163} The idea of DNP is to transfer equilibrium polarization of electron spins to nuclear spins by pumping Electron Paramagnetic Resonance (EPR) transitions of stable paramagnetic compounds added to the sample, i.e., ``polarizing agents''. Ideally, the NMR signal enhancements are reaching a value equivalent to the ratio $\gamma_e/\gamma_N$ (with $\gamma_e$ and $\gamma_N$ being the electron and nuclear gyromagnetic ratios), which is equal to 660 for protons, i.e., for nuclei with the highest $\gamma$-ratio. However, achieving the maximal theoretically allowed enhancement (or even approaching it) is still a big challenge, notably, in high magnetic fields where the polarization transfer efficiency is expected to decrease.

Historically, the first DNP mechanism to be discovered was the Overhauser mechanism,\cite{Overhauser1953} relying on electron-nuclear cross-relaxation and thus requiring fluctuations of the electron-nuclear hyperfine coupling (HFC). In insulating solids, the Overhauser mechanism was deemed to be inefficient. For this reason, solid-state DNP has relied on other  mechanisms, known as solid-effect,\cite{Jeffries57,HOVAV2010} cross-effect\cite{Hwang67,Jeffries57,HOVAV2012}  and thermal mixing.\cite{Abragam_1978} However, quite surprisingly, in some cases the Overhauser effect is operative in insulating solids,\cite{Can14} moreover, the enhancement scales favorably with the magnetic field (increasing upon the field increase from 9.4 to 18.8 Tesla).\cite{Lelli15} It is worth noting that in the case of Overhauser DNP, microwave pumping is performed on allowed NMR transitions, which are easier to saturate, providing a possible solution to the problem of limited microwave power available at high frequencies. Since Overhauser DNP in solids is an efficient mechanism at high magnetic fields (used to improve the NMR resolution) investigation of this phenomenon and further optimization of the enhancement is of great interest.

So far, Overhauser DNP in insulating solids has been reported for a single specific polarizing agent -- the BDPA radical.\cite{Can14,Lelli15} \emph{Ab initio} electronic structure calculations have shown that this radical is a mixed-valence compound\cite{Heckmann} in which the electron spin density is spontaneously hopping between the two sites, giving rise to fluctuations of the HFC and, consequently, to cross-relaxation.\cite{BDPA_DMRG, PylaevaJPCL17} According to molecular dynamics simulations the spectral density of the fluctuations is peaking at frequencies around 100-700 GHz, providing favorable conditions for electron-nuclear cross-relaxation at high fields and thus to DNP enhancements.\cite{PylaevaJPCL17} Interestingly, a similar mixed-valence behaviour was reported even before the BDPA in a flavin derivative.\cite{Flavin}

\begin{figure}
\includegraphics[width=0.9\columnwidth]{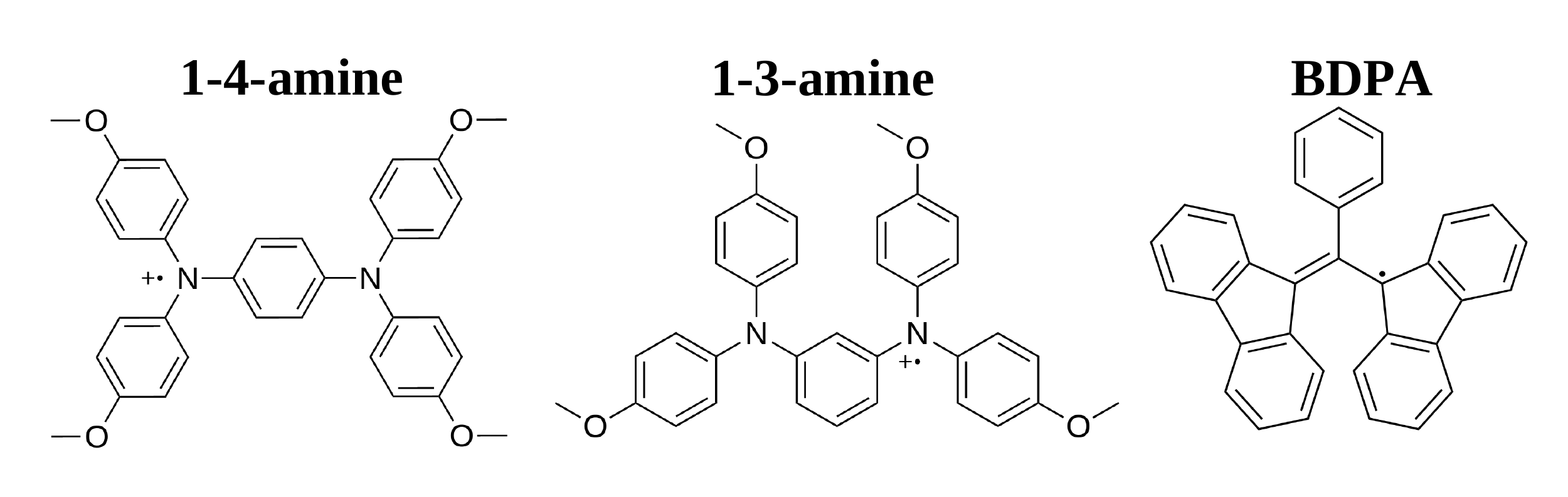}
  \caption{Set of the radicals investigated: 1,1,4,4-tetrakis(4-methoxyphenyl)benzene-1,4-diamine (1-4-amine), 1,1,4,4-tetrakis(4-methoxyphenyl)benzene-1,3-diamine (1-3-amine), and 1,3-bis(diphenylene)-2-phenylallyl (BDPA) radical}
  \label{fig:structures}
\end{figure}

Presently, experimental data on Overhauser DNP in insulating solids are limited, as the effect has been reported for a single polarizing agent. The goal of this work is thus (i) to verify the theoretical prediction that mixed-valence compounds are suitable polarizing agents for Overhauser DNP and potentially (ii) improving the enhancement provided by Overhauser DNP in solids at high magnetic fields. To this end, we here study a new set of mixed-valence radicals, namely 1,1,4,4-tetrakis(4-methoxyphenyl)benzene-1,4-diamine (1-4-amine) radical, and 1,1,4,4-tetrakis(4-methoxyphenyl)benzene-1,3-diamine (1-3-amine) radical shown in Figure \ref{fig:structures} and compare their performance in DNP experiments to those of the 1,3-Bis(diphenylene)-2-phenylallyl (BDPA) radical. Based on our previous work, these radicals were chosen from a theoretical screening of a larger number of potential mixed-valence candidates using electronic structure calculations. The most promising candidates were synthesized, and we here present the data of an EPR study of these new radicals and also the results of DNP experiments performed at the magnetic field of 18.8 Tesla. Our results indeed show that mixed-valence compounds are suitable polarizing agents for DNP and support the idea that Overhauser DNP in solids is due to the transitions between "alternative" valence structures of such compounds, which give rise to the required fluctuations of HFCs.

{\bf Methods}

{\bf Computational details}. DFT calculations were performed using Gaussian16.\cite{g16} Geometry optimization was performed at the BMK/TZVPP level,\cite{BMK,Weigend2006} followed by calculations of magnetic properties at the BMK/EPR-III level;\cite{BaroneEPR} ultrafine convergence and integral treatment was employed in all calculations. CASSCF calculations were done in ORCA at the CASSCF(3,3)/def2-TZVP level of theory.\cite{NeeseORCA, Weigend2006}

{\bf Experimental details}. Synthesis and sample preparation details are given in Supporting Information. Echo-detected EPR spectra were obtained using a home built G-band EPR spectrometer (180 GHz, 6.4 T).\cite{rohrer2001high} EPR spectra were recorded in a 1,1,2,2-tetrachlorethane (TCE) matrix with a radical concentration of ~ 0.1 - 0.5 mM. G-band echo detected EPR spectra were recorded at 50 K using a pulse length of 44 ns and 70 ns for $\pi$/2 and $\pi$ pulses, respectively; the inter-pulse delay was 200 ns. To determine the values of the $g$-factor of nitroxides, we placed a $^{55}$Mn$^{2+}$ standard sample (g(Mn${^{2+}}$)=2.00101) in the resonator together with the studied sample. The $g$-tensor parameters of all radicals under study were obtained from simulation of the G-band EPR spectra with the EasySpin program\cite{Easyspin} using function pepper, in corresponding solid-state regime.

DNP experiments were on a 800 MHz / 527 GHz NMR/DNP spectrometer (Bruker BioSpin) equipped with a sweep coil that allowed to vary the B$_{0}$ magnetic field in the range of $\pm$45 mT. The MAS frequency was set to 8 kHz unless stated otherwise. The DNP enhancement was obtained by comparing the ${^1}$H signals of TCE with and without MW irradiation using a rotor-synchronized Hahn echo pulse sequence after a series of saturation pulses.

\begin{figure}
\includegraphics[width=0.9\columnwidth]{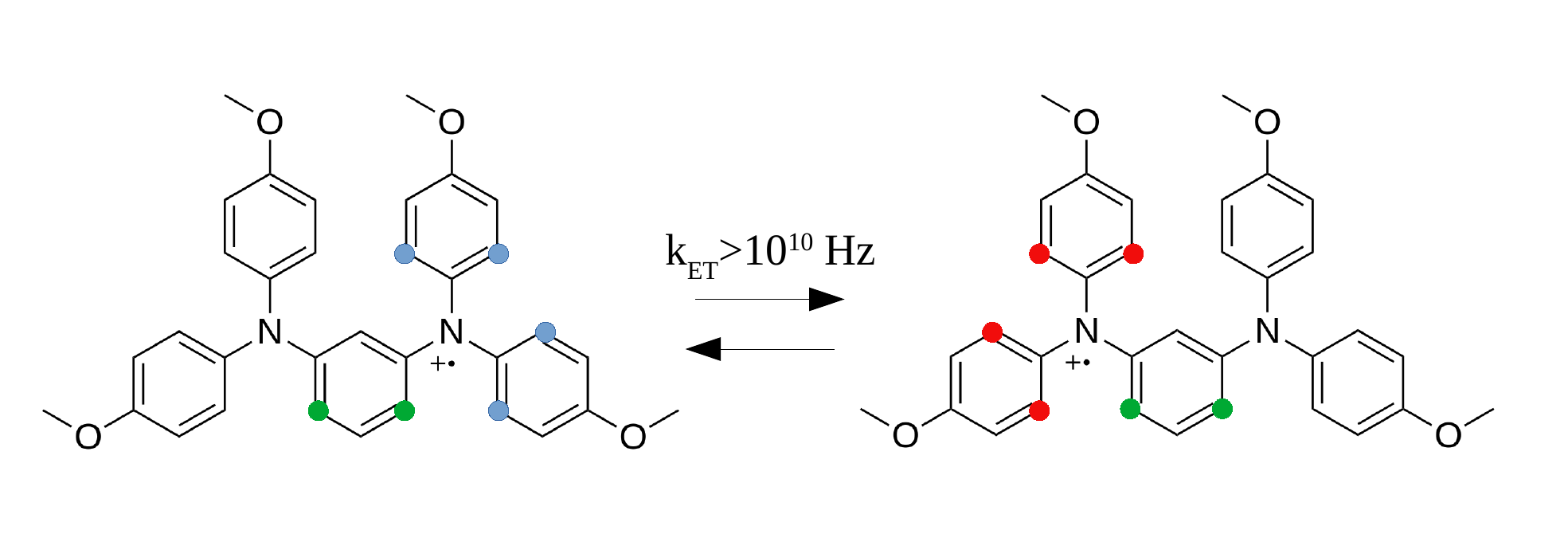}
  \caption{Changes in hyperfine coupling pattern due to electron transfer in 1-3-amine (BMK/EPR-III calculation). Rate of the electron transfer was estimated by Uebe et al.\cite{Uebe19}}
  \label{fig:hfcs}
\end{figure}

{\bf Results and Discussion}
Compounds like BDPA are known as mixed-valence compounds. Such systems are also called
(pseudo) Jahn-Teller systems,\cite{Heckmann, BersukerPJT} where electronic and vibrational degrees of freedom are coupled. According to a classification scheme by Robin and Day,\cite{Robin1968} BDPA belongs specifically to the class II mixed-valence compound.\cite{BDPA_DMRG}  Such compounds have a localized electronic state with a barrier in the center. Accordingly, the two valence states exhibit a coupling interaction of intermediate strength. The interaction is sufficiently weak to prevent a collapse into one symmetric state but strong enough to reduce the height of the energy barrier in the middle. In general, the interconversion of two structures occurs upon excitation: thermally, when higher vibrational states are populated, or via tunneling through the barrier.\cite{MazurIR,MazurESR} It is worth mentioning that changes of HFC pattern with temperature are often used to estimate the electron transfer rate in mixed-valence compounds.\cite{Heckmann, Uebe19} Recently, some of us have confirmed that BDPA belongs to class II mixed-valence compounds using high level electronic structure methods.\cite{BDPA_DMRG}
Novel radicals have been chosen based on intensive literature search with a few considerations in mind, i.e., fast electron transfer rate and narrow EPR line.\cite{Heckmann,wengerRev,KauppRev} Both 1-3-amine and 1-4-amine have been reported to have a mixed valence character close to class II/III border.\cite{Heckmann, LambertJPCA04,Uebe19} Calculations of the $g$-tensor revealed that both radicals have relatively narrow EPR lines at high fields (g$_{aniso}$(1-3-amine) = 1193.8 ppm, g$_{aniso}$(1-4-amine) = 1432.9 ppm, g$_{aniso}$(BDPA) = 575.7 ppm). In 1-3-amine spin density is mostly localized on one side of the molecule which is indicated by values of the isotropic hyperfine coupling constants (see Figure~\ref{fig:hfcs}, full list of HFC constants is provided in the Supporting Information). DFT calculations of 1-4-amine in vacuum yielded class III structure where spin density is delocalized over the entire molecule.\cite{Kaupp09} However, preliminary CASSCF calculations point on a class II structure. Furthermore, solvent and counter-ion effects are known to influence class of mixed valence compounds: polar solvents as well as more compact counter-ions tend to stabilize localized class II structures. 

\begin{figure}
\includegraphics[width=0.9\columnwidth]{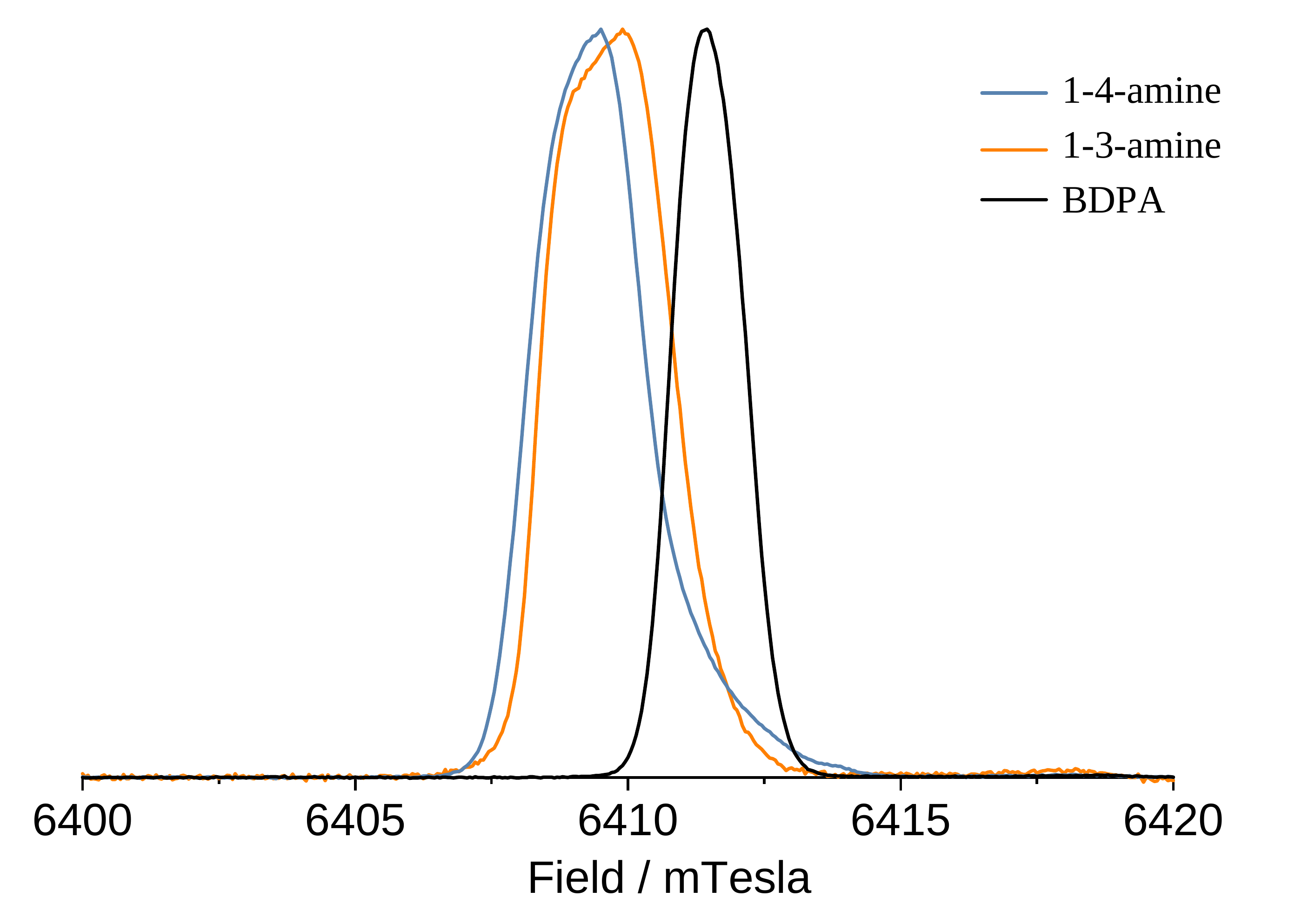}
  \caption{EPR spectra of the solutions of radicals in TCE at 50 K, acquired at 180 GHz EPR spectrometer.}
  \label{fig:EPR}
\end{figure}

EPR spectra of the compounds under study taken at 6.4 Tesla are shown in Figure \ref{fig:EPR}, along with the spectrum of BDPA shown for comparison. Both 1-3-amine and 1-4-amine exhibit a narrow EPR line corresponding to $g=2.00359$, which is inhomogeneously broadened due to $g$-tensor anisotropy. Compared to BDPA, the resonance is found at a lower field, and the EPR line is almost two times broader. However, the EPR linewidth (when measured in frequency units) is still smaller than the NMR frequency $\omega_N$ at the same field. Consequently, in DNP experiments one should expect clearly resolved contributions from the solid-effect and Overhauser effect, as previously observed for BDPA. \cite{Can14}

\begin{figure}
\includegraphics[width=0.9\columnwidth]{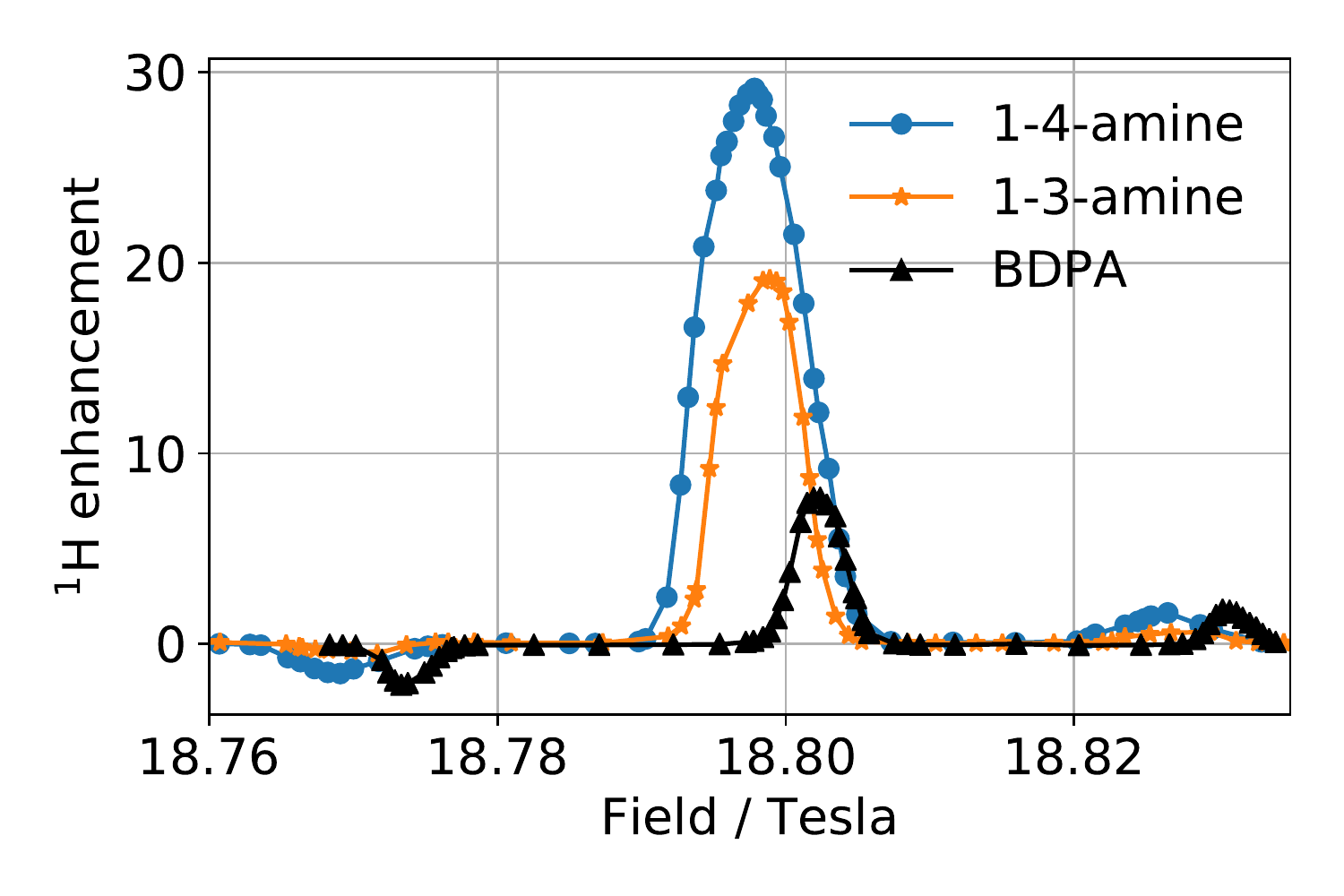}
  \caption{Field profile for the set of radicals measured at 100 K in TCE (90D:10H) matrix under 8 kHz MAS. Sample preparation details are given in the SI}
  \label{fig:profile}
\end{figure}

Figure \ref{fig:profile} shows the DNP enhancement measured at a constant microwave frequency of 527 GHz as a function of the external magnetic field, which was incremented in small steps between subsequent DNP measurements. In accordance with our expectations, individual components are clearly resolved in each spectrum, with the outer components (negative component at lower field and positive component at higher field) corresponding to the solid effect and the central component corresponding to Overhauser DNP, in qualitative agreement with studies on BDPA reported before.\cite{Can14} However, the newly proposed radicals show a better performance not only in terms of the maximum enhancement (found for the central component corresponding to Overhauser DNP) but also in terms of the ratio of the enhancement determined for the central component and outer components (which stands for the relative efficiency of Overhauser DNP and solid-effect DNP). The maximal enhancement factor found here reaches approximately 30 for 1-4-amine and 20 for 1-3-amine, whereas we found for BDPA a maximum enhancement of 8 under the same conditions.

\begin{figure}
\includegraphics[width=0.48\columnwidth]{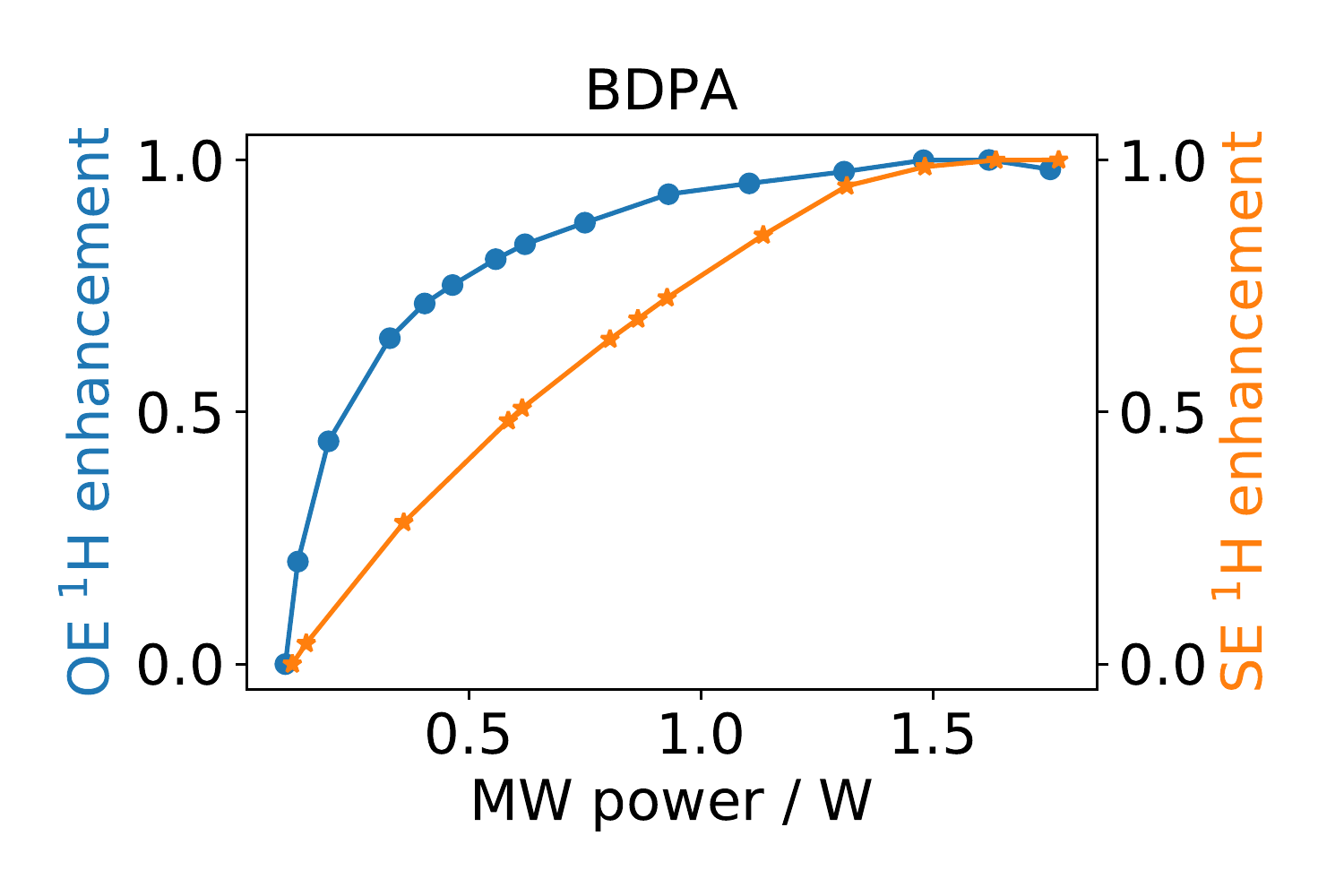}
\includegraphics[width=0.48\columnwidth]{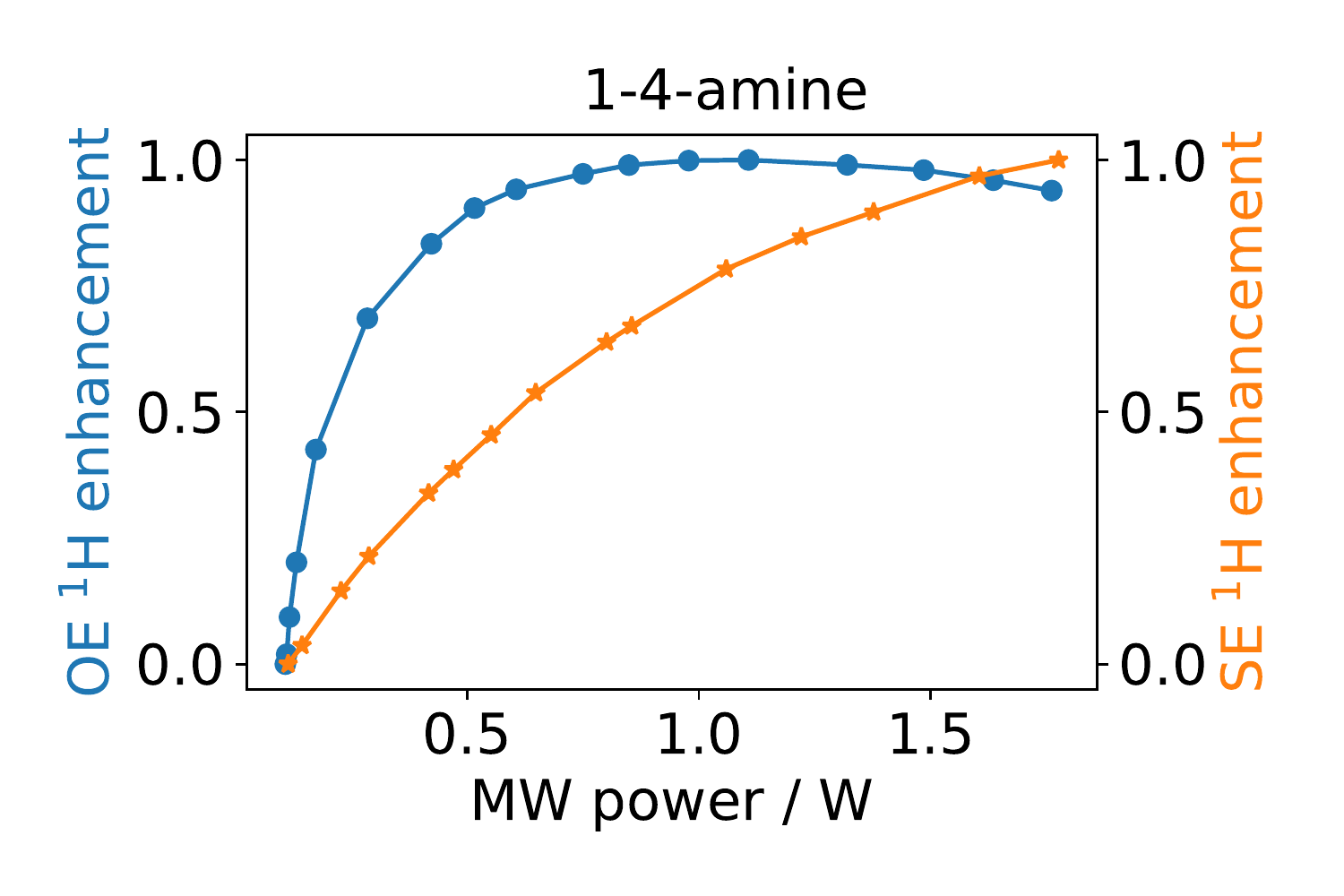}
  \caption{Normalized $^1$H DNP enhancement as a function of MW power for different components of the field profile: for BDPA (left) and 1-4-amine (right). Data for 1-3-amine is shown in Figure S2 in the SI.}
  \label{fig:MWprofile}
\end{figure}

\begin{figure}
\includegraphics[width=0.9\columnwidth]{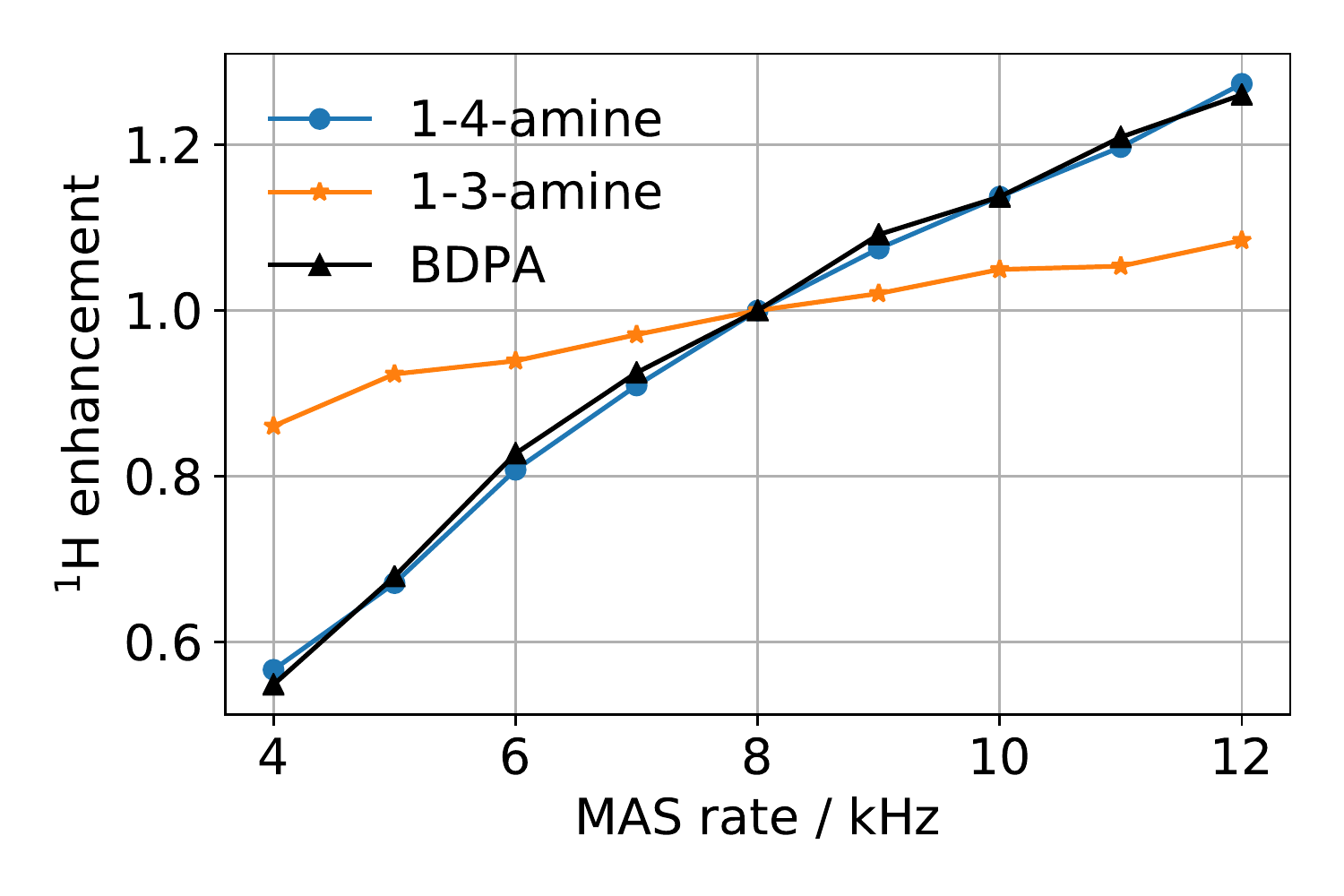}
  \caption{Normalized enhancement as a function of MAS frequency for the studied radicals..}
  \label{fig:MAS}
\end{figure}

To gain additional insight into DNP process we have also measured the dependence of the enhancement on the pumping power for the central component and one of the outer components. Such dependencies are expected to be different \cite{Can14} because in the former case pumping is performed on the ''allowed'' EPR transition, whereas in the latter case ``forbidden'' EPR transitions are irradiated. Hence, different transitions are expected to be saturated at different microwave power. As shown in Figure~\ref{fig:MWprofile} this is indeed the case for both radicals under study. Notably, for the central transition the enhancement approaches its maximal value at the power of 0.6 Watt, whereas for the forbidden transitions the enhancement keeps increasing at the highest available power. The enhancement coming from Overhauser effect thus scales favorably with microwave power, in contrast to solid-effect DNP.
 Lastly, we also investigated the MAS dependence (Figure~\ref{fig:MAS}) for the new radicals which showed a similar behavior to previous OE DNP studies.\cite{fastMAS_oe}

In summary, we have experimentally observed novel mixed-valence radicals inducing hyperpolarization based on the Overhauser mechanism in insulating solids. Our findings are based on a set of complementary methods: computer simulations, high field EPR and DNP measurements. We plan to continue our investigation of mixed-valence polarizing agents with special focus on their stability under physiological (i.e., aqueous solution, higher temperature) conditions both experimentally and theoretically.

\begin{acknowledgement}
This work has been supported by iNEXT-Discovery, grant number 871037, funded by the Horizon 2020 program of the European Commission.
Svetlana Pylaeva would like to thank Prof. Matvey Fedin, Dr. Yury Lebedev for fruitful discussions of EPR experiments and organic synthesis. 
HE acknowledges the DFG for funding his temporary position as a principal investigator (Eigene Stelle). 
TDK acknowledges funding from the European Research Council (ERC) under the European Union’s Horizon 2020 research and innovation program (grant agreement no. 716142).
The generous allocation of supercomputer time by the Paderborn Center for Parallel Computing (PC2) is kindly acknowledged.

JP acknowledges the DFG for funding (PA1562-14-1).

The DNP experiments were supported by the Dutch Research Council (NWO grants 700.11.344 and 700.58.102 to MB).
\end{acknowledgement}

\begin{suppinfo}
Synthesis, sample preparation procedures, simulations' results are described in the supporting information. 
\end{suppinfo}

\bibliography{bib}

\end{document}